\documentclass[preprint,prb]{revtex4}

\usepackage{graphicx}
\usepackage{DColumn}
\usepackage{float}
\usepackage{amsmath}

\newcommand{\comment}[1]{}

\begin{document}

\title{\boldmath Hidden non-Fermi liquid behavior caused by magnetic phase transition in Ni-doped Ba-122 pnictides. \unboldmath}


\author{Seokbae Lee$^1$, Kiyoung Choi$^{2}$, Eilho Jung$^{1}$, Seulki Rho$^{1}$, Soohyeon Shin$^{1}$, Tuson Park$^{1}$ \& Jungseek Hwang$^{1,*}$}

\affiliation{$^{1}$Department of Physics, Sungkyunkwan University, Suwon, Gyeonggi-do 440-746, Republic of Korea}
\affiliation{$^{2}$Center for Novel States of Complex Materials Research, Department of Physics and Astronomy, Seoul National University, Seoul 151-747, Republic of Korea}

\date{\today}

\begin{abstract}

{\bf We studied two BaFe$_{2-x}$Ni$_x$As$_2$ (Ni-doped Ba-122) single crystals at two different doping levels (underdoped and optimally doped) using an optical spectroscopic technique. The underdoped sample shows a magnetic phase transition around 80 K. We analyze the data with a Drude-Lorentz model with two Drude components ($D1$ and $D2$). It is known that the narrow $D1$ component originates from electron carriers in the electron-pockets and the broad $D2$ mode is from hole carriers in the hole-pockets. While the plasma frequencies of both Drude components and the static scattering rate of the broad $D2$ component show negligible temperature dependencies, the static scattering rate of the $D1$ mode shows strong temperature dependence for the both samples. We observed a hidden quasi-linear temperature dependence in the scattering rate of the $D1$ mode above and below the magnetic transition temperature while in the optimally doped sample the scattering rate shows a more quadratic temperature dependence. The hidden non-Fermi liquid behavior in the underdoped sample seems to be related to the magnetic phase of the material.}
\\ \\

\noindent *Correspondence to [email: jungseek@skku.edu].

\end{abstract}


\maketitle

The FeAs high temperature superconductors have been attracted much attention and have been studied intensively since their first discovery in 2006 and 2008\cite{kamihara:2006,kamihara:2008}. These compounds are of several different types. Those types are a "1111" type (LaFePO, SmFeAsO, PrFeAsO, LaFeAsO etc.), a "122" type (BaFe$_2$As$_2$, SrFe$_2$As$_2$, CaFe$_2$As$_2$ etc.), a "111" type (LiFeAs, NaFeAs, LiFeP etc.), a "11" type (Fe(Te,Se)), and so on. Each type except "11" has a different layered structure consisting of an alternating a charge transport layer and a charge reservoir layer; "11" type consists of only charge transfer layers. Each type shows a similar phase diagram as that of cuprates \cite{basov:2011} even though the details are quite different; undoped parent compounds are antiferromagnetic metals while cuprates are antiferromagnetic Mott insulators. The superconducting mechanism in this material has not yet been figured out. Researchers in this field believe that its mechanism may have a common origin with that of the cuprates. Density functional theory calculation shows the electronic structure of FeAs compounds is multiband with three hole-like bands at the $\Gamma$ point and two electron-like bands at the Brillouin zone corners\cite{subedi:2008}. The FeAs-compounds also show multigap superconductivity\cite{ding:2008} compared with the cuprates which have a single $d$-wave superconducting gap. Since this material has the multiband channels an analysis of the optical data may need to include two different types of free-carrier contributions\cite{wu:2010}. This material also shows quite strong correlations among charge carriers; electronic many-body effects seem to be important\cite{qazilbash:2009}.

Recent optical study on Ba$_{0.6}$K$_{0.4}$Fe$_2$As$_2$ with two Drude modes shows interesting hidden temperature-dependent properties of the two Drude modes\cite{dai:2013}. It is observed that the two Drude modes show significantly different temperature-dependent trends; only one of the Drude modes shows a strong temperature dependence. The approach used is a similar to the so-called {\it two-component} analysis introduced in an earlier paper on Bi$_2$Sr$_2$CaCu$_2$O$_{8+\delta}$ cuprates\cite{quijada:1999}. In the earlier paper they also introduced a so-called {\it one-component} analysis, which is also known as an extended Drude model\cite{puchkov:1996,hwang:2004}. Since FeAs-compounds have multibands the compounds can be analyzed approximately with two Drude components as in Dai {\it et al.} paper\cite{dai:2013}, which corresponds to the {\it two-component} analysis for cuprates. The {\it one-component} (or extended Drude model) analysis has been applied to Fe-pnictide systems\cite{yang:2009a,wu:2010b,hwang:2015} with a single band approximation even though the material systems have multiband characteristics. To resolve the multiband issue one should develop a new method including the multiband nature. Another issue for the direct application of the extended Drude model to the Fe-pictides is that the system has low-energy interband transitions\cite{benfatto:2011}, which need to be considered. Because of those nontrivial issues associated with application of the one-component analysis we applied a two-component analysis to analyze our optically measured spectra.

In this paper we investigate Ni-doped Ba-122 type Fe-pnictide samples (BaNi$_x$Fe$_{2-x}$As$_2$) at two different doping levels: an underdoped $x$ = 0.05 with the superconducting transition temperature, $T_c = 10$ K and an optimally doped $x$ = 0.10 with $T_c = 17$ K. The underdoped compound shows magnetic and structure transitions near 80 K. We applied the {\it two-component} analysis to understand our optical data of BaFe$_{2-x}$Ni$_x$As$_2$ ($x =$ 0.05 and 0.10). For the analysis we use a model with two Drude components for the two electron- and hole- pockets on the Fermi surface. We found that one of the two Drude modes is quite narrow compared to the other as in a reported paper\cite{dai:2013}. We denote the narrow as $D1$ and the broad $D2$. Other studies show that $D1$ ($D2$) come from the contribution of the electron-pocket (hole-pocket)\cite{fang:2009,shen:2011,dai:2013}. From this analysis we can expose hidden transport properties. We found that in the underdoped sample the magnetic transition affects transport properties of electron carriers. We observed that the underdoped and optimally doped samples show different temperature dependent hidden transport properties; the optimally doped sample shows a Fermi-liquid behavior while the underdoped one shows a non-Fermi liquid behavior.

\section*{Results and discussion}

The measured reflectance spectra of our two Ni-doped Ba-122 crystal samples (see Methods section) at various temperatures are shown in Fig. 1 (a, b). In the insets we display magnified views to show the data better in low frequency region. While the optimally doped sample ($x =$ 0.10) shows a monotonic temperature evolution the underdoped one ($x =$ 0.05) shows non-monotonic temperature evolution below $\sim$ 1000 cm$^{-1}$; reflectance increases initially as temperature decreases down to $\sim$ 100 K and then below this temperature it decreases as shown in the inset of the upper panel. The non-monotonic behavior with temperature seems to be related to the magnetic phase transition of the system. Both sets of data show two characteristic crossing points near 1000 cm$^{-1}$ and 4000 cm$^{-1}$. We analyzed further the reflectance spectra using a Kramers-Kronig relation (see Methods section) to obtain optical constants including the optical conductivity and the dynamic dielectric function.

In Fig. 2 (a, b) we display the real part of the optical conductivity ($\sigma_1(\omega)$) of our two samples obtained using the Kramers-Kronig process described in Methods section. As we expected from the reflectance data we can see a conductivity inversion at low frequency in the conductivity of the underdoped sample while the optimally doped sample shows a monotonic increase as temperature decreases. We also display the measured DC resistivity data obtained by using a four-probe measurement technique in the insets. The DC resistivity of the underdoped sample shows an anomaly near 80 K marked with a red arrow as shown in the inset of the upper panel where the magnetic phase transition takes place. To study the frequency dependent spectral weight redistribution we calculated the accumulated conductivity (or partial sum), which is defined as $W_s(\omega) \equiv \int_{0}^{\omega}\sigma_1(\omega') d\omega'$. We note that $W_s(\infty)$ is proportional to the number density of total electrons in a material, {\it i.e.} $8\:W_s(\infty) \equiv \sum_i 4\pi N_i e^2/m_i^*$ where $N_i$ is the electron density of the $i$th band, $e$ is the unit charge, and $m_i^*$ is the effective mass (or band mass) of electron in the $i$th band. We display the partial sums of our two samples at various temperatures in Fig. 3 (a, b). We can see spectral weight redistributions as a function of temperature. Our optimally doped sample (in Fig. 3b) shows monotonic temperature dependence; as temperature decreases the spectral weight increases in the low frequency region and then the spectral weights at all measured temperatures merge together at high frequency. We display the optical conductivity in low frequency region below 700 cm$^{-1}$ in the inset to show the temperature dependent behavior more clearly. Our underdoped sample (in Fig. 3a) shows a dramatic change across the magnetic transition temperature; below the transition temperature a significant spectral weight loss in low frequency region is observed and the spectral weight loss seems to be transferred to high frequency region above our measurement range. We also display the optical conductivity in the low frequency region below 700 cm$^{-1}$ in the inset to show the non-monotonic temperature dependent behavior in the spectral weight.

We display two representative real parts of the optical conductivity data of our two samples at 150 K and their Drude-Lorentz fits (see Methods section) in Fig. 4 (a, b). For the fitting we used two Drude modes ($D1$: one narrow and $D2$: the other broad) and two Lorentz modes ($L1$: one is strong at high and $L2$: the other weak at low frequency). While the narrow Drude component ($D1$) is known as contribution from electron carriers in the electron-pocket, the broad one ($D2$) from hole carriers\cite{fang:2009,shen:2011,dai:2013}. The Lorentz component ($L1$) at high frequency shows negligible temperature dependence. The Lorentz component ($L2$) at low frequency is the interband transition at 1300 cm$^{-1}$ which was reported recently by Marsik {\it et.al}\cite{marsik:2013}. For the fitting we used the reported temperature and doping dependent amplitudes of the low-energy Lorentz component. We note that the low-energy component is quite weak in terms of the spectral weight. Therefore we expect that it will have negligible impact on overall fits. We display temperature dependent fitting parameters of the two Drude components of the underdoped and optimally doped samples in Fig. 5 (a, b) and Fig. 6 (a, b), respectively.

In Fig. 5 (a-d) we show four quantities of two Drude modes ($D1$ and $D2$) in the underdoped sample at various temperatures below and above the magnetic transition temperature ($\sim$ 80 K) obtained from the fitting. The four quantities are the plasma frequency ($\Omega_{Di,p}$), the static scattering rate ($1/\tau_{Di}$), the DC conductivity ($\sigma_{dc}$), and the DC resistivity ($\rho$) of the two Drude modes. Both plasma frequencies (in panel (a)) show negligible temperature dependencies. While the static scattering rate of the $D1$ Drude mode shows strong temperature dependence that of the $D2$ mode shows negligible temperature dependence (in panel (b)). These are similar behaviors to what was observed by Dai {\it et al.}\cite{dai:2013}; only the scattering rate of the $D1$ shows significant temperature dependence. We note that their sample was an optimally hole-doped (or K-doped) Ba-122 sample. Here we observed that our underdoped sample shows an interesting upturn below the magnetic transition temperature (marked with a red arrow), which is absent in both optimally hole-doped Ba$_{0.6}$K$_{0.4}$Fe$_2$As$_2$ sample in Dai {\it et al.} paper\cite{dai:2013} and our optimally electron-doped one. As shown in panel (b) the static scattering rate of the $D1$ mode decreases linearly down to the magnetic transition temperature from 300 K and then it increases linearly below this temperature. This linear temperature dependent behavior indicates that the hidden transport property is the non-Fermi-liquid above and below the transition temperature. The linear temperature dependent behavior seems to be similar to that in non-Fermi liquid phase of cuprates\cite{ito:1993}. Here we point out that the magnetic phase transition in the underdoped sample is closely associated with the $D1$ mode or electron carriers in the electron-pocket. It seems to be consistent with earlier angle-resolved photoemission study of detwinned single crystals of electron-doped Ba(Fe$_{1-x}$Co$_x$)$_2$As$_2$\cite{yi:2011}. We observe similar temperature dependent non-trivial behavior in the other two quantities [$\sigma_{dc}(T) \equiv \Omega_p(T)^2/[4\pi\cdot\:1/\tau_D(T)]$ and $\rho(T) \equiv 1/\sigma_{dc}(T)$] shown in the right panels (c, d). In panel (d) we also display the total DC resistivity (open star symbols) which can be calculated using a relation, $1/\rho_{D1+D2} = 1/\rho_{D1}+1/\rho_{D2}$.

In Fig. 6 (a-d) we display the four quantities of our optimally doped sample at various temperatures. We can see that only the static scattering rate of $D1$ shows significant temperature dependence as we have seen previously in the underdoped sample. In panel (a) we observe that the plasma frequency of the $D2$ mode (the hole carrier density) is reduced and that of the $D1$ (the electron carrier density) is slightly enhanced as we expected; the system is more electron-doped. In panel (b) we do not observe the non-monotonic and non-Fermi liquid behavior in the static scattering rate of $D1$ mode which we observed in the underdoped sample (see Fig. 5b). The $D1$ scattering rate decreases monotonically as temperature decreases. We do not see the $T$ linear dependence which was observed in the optimally doped Ba$_{0.6}$K$_{0.4}$Fe$_2$As$_2$ by Dai {\it et al.}\cite{dai:2013} instead our data show a rapid decrease down to 200 K and then a less rapid decrease below this temperature. The different hidden temperature dependent trends might be related to different temperature dependencies in the measured DC resistivity data of two materials with different doping types: one in the inset of Fig. 2b and the other in the inset of Fig. 1 of Ref. \cite{dai:2013}. In the inset we display the $D1$ scattering rate as a function of $T^2$; its temperature dependence is near quadratic, which is similar to that of near optimally electron-doped (Co- and Ni-doped) samples reported by Barisic {\it et al.}\cite{barisic:2010}. Here we note that the optimally electron-doped sample seems to show different hidden transport property from that of the hole-doped sample in Ba-122 material Fe-pnictide systems. We also display $\sigma_{dc}(T)$ and $\rho(T)$ in Fig. 6 (c) and (d), respectively. In panel (d) we show the total DC resistivity (solid star symbols) to compare it with the measured DC resistivity in the inset of Fig. 2b; even though the two data sets show similar temperature dependence they also show different residual resistivities, which we do not know clearly yet.

\section*{Conclusion}

We studied two different Ni-doped (or electron-doped) Ba-122 single crystal samples, BaFe$_{2-x}$Ni$_x$As$_2$ ($x $= 0.05: underdoped and $x $= 0.10: optimally doped), using an optical spectroscopic technique. The DC resistivity of our underdoped sample shows a magnetic phase transition near 80 K. We applied the so-called {\it two-component} approach\cite{quijada:1999}, which is also called the Drude-Lorentz analysis, to disclose their hidden transport properties. The underdoped sample shows significantly different transport properties below and above the transition temperature. Above the transition temperature the static scattering of the $D1$ mode (or the narrow mode) shows a quasi-linear temperature behavior which is different from that of the optimally doped sample ($x=$ 0.10). The optimally doped sample shows near Fermi-liquid (quadratic temperature dependent) behavior in the measured temperature range. Below the magnetic transition temperature the underdoped sample shows completely different transport properties; we observed an increase linear law on lowering the temperature for the static scattering rate of the $D1$ Drude mode (see Fig. 5b). We also found that the magnetic phase transition influences only electron carriers of the electron-pocket on the Fermi surface; as the system enters into the ordered magnetic phase the $D1$ static scattering increases linearly with temperature. The holes in the hole-pocket seems not to be affected by the magnetic ordering in the system. The scattering rate of $D1$ mode in the optimally doped sample shows near quadratic temperature dependence, which is consistent with those of near optimally electron-doped samples reported by Barisic {\it et al.}\cite{barisic:2010}. However, this temperature dependent transport behavior is different from that of optimally hole-doped sample investigated by Dai {\it et al.}\cite{dai:2013}. This indicates that the electron- and hole-doped samples show different hidden $D1$ transport properties. We believe that these new findings may help to figuring out the superconducting mechanism of relatively new high temperature superconductors, Fe-pnictides.

\newpage

\begin{figure}[!htbp]
  \vspace*{-0.3 cm}%
  \vspace*{-0.7 cm}%
\caption{(a, b) The measured reflectance spectra of BaFe$_{2-x}$Ni$_x$As$_2$ ($x =$ 0.05 and 0.10) at various temperatures.
In the insets we display magnified views of reflectance spectra at low frequency region.}
\end{figure}

\begin{figure}[!htbp]
  \vspace*{-0.3 cm}%
  \vspace*{-0.7 cm}%
\caption{(a, b) The extracted conductivity spectra of BaFe$_{2-x}$Ni$_x$As$_2$ ($x =$ 0.05 and 0.10)
at various temperatures using the Kramers-Kronig analysis. In the insets we display measured DC resistivity
data. The red arrow in the inset of the upper panel indicates the onset temperature of the magnetic phase
transition.}
\end{figure}

\begin{figure}[!htbp]
  \vspace*{-0.3 cm}%
  \vspace*{-0.7 cm}%
\caption{(a, b) The partial sums of BaFe$_{2-x}$Ni$_x$As$_2$ ($x =$ 0.05 and 0.10) at various temperatures.
In the insets we display magnified views of the optical conductivity spectra of the two samples
at low frequency region below 700 cm$^{-1}$.}
\end{figure}

\begin{figure}[!htbp]
  \vspace*{-0.3 cm}%
  \vspace*{-0.7 cm}%
\caption{(a, b) Two representative Drude-Lorentz fits and optical conductivity data of BaFe$_{2-x}$Ni$_x$As$_2$ ($x =$ 0.05 and 0.10) at $T =$ 150 K. We used two Drude components ($D1$ and $D2$) and also included a low-energy interband transition observed
by Marsik {\it et al}\cite{marsik:2013}.}
\end{figure}

\begin{figure}[!htbp]
  \vspace*{-0.3 cm}%
  \vspace*{-0.2 cm}%
\caption{(a, b) The fitting parameters ($\Omega_{Di,p}$ and $1/\tau_{Di}$) of two Drude components ($i =$ 1 and 2) of 
BaFe$_{1.95}$Ni$_{0.05}$As$_2$. (c, d) We also display the calculated DC conductivities ($\sigma_{dc}(T)$) and DC resistivity 
data ($\rho(T)$) including the total resistivity as functions of temperature, $T$.}
\end{figure}

\begin{figure}[!htbp]
  \vspace*{-0.3 cm}%
  \vspace*{-0.2 cm}%
\caption{(a, b) The fitting parameters ($\Omega_{Di,p}$ and $1/\tau_{Di}$) of two Drude components ($i =$ 1, 2) of BaFe$_{1.90}$Ni$_{0.10}$As$_2$. In the inset we display the static scattering rate of $D1$ mode as a function of $T^2$. (c, d) We also display the calculated DC conductivities ($\sigma_{dc}(T)$) and DC
resistivity data($\rho(T)$) including the total resistivity as functions of temperature, $T$.}
\end{figure}

\newpage

\section*{Methods}

\subsection*{Samples and reflectance measurement technique.}
High quality single crystal BaNi$_x$Fe$_{2-x}$As$_2$ ($x =$ 0.05 and 0.10) samples were grown using a high-temperature self-flux method\cite{canfield:1992,wang:2009,ni:2009}. The areas of samples are roughly 2$\times$2 mm$^2$ with a thickness $\leq$ 200 $\mu$m. We had to add a thin metal plate between the sample and a sample cone to support and prevent bending of the sample caused by an epoxy contraction when the temperature was reduced. The transition temperatures were determined using magnetic susceptibility and DC-transport measurements. The DC transport resistivity data of two samples are shown in insets of Fig. 1. We obtained accurate reflectance spectra of our samples at various temperatures using an in-situ metallization method\cite{homes:1993} and a continuous liquid Helium flow system. We used a commercial FTIR-type spectrometer, Bruker Vertex 80v to take the reflectance spectra in far- and mid-infrared range (80 - 8000 cm$^{-1}$ or 10 meV - 1.0 eV).

\subsection*{Kramers-Kronig analysis.}
To get the optical constants from the measured reflectance spectra we performed a Kramers-Kronig analysis method\cite{wooten}. To do the analysis we have to have data in a complete frequency range from zero to infinity. Practically we can have measured data only in a finite spectral range because of experimental limitations. We need to extrapolate the measured data to both directions: zero and infinity. We used the well-known Hagen-Rubens relation for the low frequency extrapolations. For the high frequency extrapolations we took advantage of existing data of similar materials (BaFe$_{1.85}$Co$_{0.15}$As$_2$)\cite{tu:2010} for the higher frequencies up to 25,000 cm$^{-1}$. Above the frequency up to 10$^6$ cm$^{-1}$ we used $R(\omega) = R(\omega_{max})\: \omega^{-DE}$, where $R(\omega)$ is reflectance, $\omega_{max}$ is the highest frequency in the measured data file, and $DE$ can be selected between 0 and 2. Above 10$^6$ cm$^{-1}$, where we are in the free electron region, we used $R(\omega) = \omega^{-P}$, where $P$ can be selected between 2 and 4\cite{wooten,tanner1}.

\subsection*{Two-component analysis: Drude-Lorentz model with two Drude modes.}
We applied the two-component analysis to the optical conductivity as in a published literature\cite{dai:2013}. We used two Drude components (one is narrow and the other is comparatively broad) and one Lorentz component and a low-energy interband transition located near 1300 cm$^{-1}$ as reported recently\cite{marsik:2013} to fit the conductivity data up to 6500 cm$^{-1}$. The complex dielectric function ($\tilde{\epsilon}(\omega) \equiv \epsilon_1(\omega)+i\epsilon_2(\omega)$) can be written in a Drude-Lorentz model as:
\begin{equation}\label{eq2}
\tilde{\epsilon}(\omega) = \epsilon_{H} - \sum_{i=1,2}\frac{\Omega_{Di,p}^2}{\omega(\omega + i \tau_{Di}^{-1})}+\sum_j\frac{\Omega_{j,p}^2}{\omega_j^2-\omega^2-i\gamma_j \omega}
\end{equation}
where $\epsilon_{H}$ is the background dielectric function, which comes from contribution of the high frequency absorption. $D1$ ($D2$) stands for a narrow Drude component (a broad Drude component), $\Omega_{Di,p}$ is the Drude plasma frequency, $\tau_{Di}^{-1}$ is the (average) elastic scattering rate among free charge carriers. $\Omega_{j, p}$, $\omega_j$, and $\gamma_j$ are the plasma frequency, the center frequency, and the width of the $j$th Lorentz component, respectively. The optical conductivity can be related to the dielectric function as $\tilde{\sigma}(\omega) \equiv \sigma_1(\omega)+i\sigma_2(\omega) = i \omega [\epsilon_{H}-\tilde{\epsilon}(\omega)]/4\pi$.


\bibliographystyle{naturemag}
\bibliography{bib}

\begin{thebibliography}{10}
\expandafter\ifx\csname url\endcsname\relax
  \def\url#1{\texttt{#1}}\fi
\expandafter\ifx\csname urlprefix\endcsname\relax\def\urlprefix{URL }\fi
\providecommand{\bibinfo}[2]{#2}
\providecommand{\eprint}[2][]{\url{#2}}

\bibitem{kamihara:2006}
\bibinfo{author}{Kamihara, Y.} \emph{et~al.}
\newblock \bibinfo{title}{Iron-based layered superconductor: \mbox{LaOFeP}}.
\newblock \emph{\bibinfo{journal}{J. Am. Chem. Soc.}}
  \textbf{\bibinfo{volume}{128}}, \bibinfo{pages}{10012}
  (\bibinfo{year}{2006}).

\bibitem{kamihara:2008}
\bibinfo{author}{Kamihara, Y.}, \bibinfo{author}{Watanabe, T.},
  \bibinfo{author}{Hirano, M.} \& \bibinfo{author}{Hosono, H.}
\newblock \bibinfo{title}{Iron-based layered superconductor
  \mbox{La[O$_{1-x}$F$_x$]FeAs} ($x$ = 0.05-0.12) with \mbox{$T_c$} = 26
  \mbox{K}}.
\newblock \emph{\bibinfo{journal}{J. Am. Chem. Soc.}}
  \textbf{\bibinfo{volume}{130}}, \bibinfo{pages}{3296} (\bibinfo{year}{2008}).

\bibitem{basov:2011}
\bibinfo{author}{Basov, D.~N.} \& \bibinfo{author}{Chubukov, A.~V.}
\newblock \bibinfo{title}{Manifesto for a higher \mbox{$T_c$}}.
\newblock \emph{\bibinfo{journal}{Nat. Phys.}} \textbf{\bibinfo{volume}{7}},
  \bibinfo{pages}{272} (\bibinfo{year}{2011}).

\bibitem{subedi:2008}
\bibinfo{author}{Subedi, A.}, \bibinfo{author}{Zhang, L.},
  \bibinfo{author}{Singh, D.~J.} \& \bibinfo{author}{Du, M.~H.}
\newblock \bibinfo{title}{Density functional study of \mbox{FeS}, \mbox{FeSe},
  and \mbox{FeTe}: Electronic structure, magnetism, phonons, and
  superconductivity}.
\newblock \emph{\bibinfo{journal}{Phys. Rev. B}} \textbf{\bibinfo{volume}{78}},
  \bibinfo{pages}{134514} (\bibinfo{year}{2008}).

\bibitem{ding:2008}
\bibinfo{author}{Ding, H.} \emph{et~al.}
\newblock \bibinfo{title}{Observation of fermi-surface-dependent nodeless
  superconducting gaps in \mbox{Ba$_{0.6}$K$_0.4$Fe$_2$As$_2$}}.
\newblock \emph{\bibinfo{journal}{Europhys. Lett.}}
  \textbf{\bibinfo{volume}{83}}, \bibinfo{pages}{47001} (\bibinfo{year}{2008}).

\bibitem{wu:2010}
\bibinfo{author}{D.Wu} \emph{et~al.}
\newblock \bibinfo{title}{Optical investigations of the normal and
  superconducting states reveal two electronic subsystems in iron pnictides}.
\newblock \emph{\bibinfo{journal}{Phys. Rev. B}} \textbf{\bibinfo{volume}{81}},
  \bibinfo{pages}{100512(R)} (\bibinfo{year}{2010}).

\bibitem{qazilbash:2009}
\bibinfo{author}{Qazilbash, M.~M.} \emph{et~al.}
\newblock \bibinfo{title}{Electronic correlations in the iron pnictides}.
\newblock \emph{\bibinfo{journal}{Nat. Phys.}} \textbf{\bibinfo{volume}{5}},
  \bibinfo{pages}{647} (\bibinfo{year}{2009}).

\bibitem{dai:2013}
\bibinfo{author}{Dai, Y.~M.} \emph{et~al.}
\newblock \bibinfo{title}{Hidden \mbox{$T$}-linear scattering rate in
  \mbox{Ba$_{0.6}$K$_{0.4}$Fe$_2$As$_2$} revealed by optical spectroscopy}.
\newblock \emph{\bibinfo{journal}{Phys. Rev. Lett.}}
  \textbf{\bibinfo{volume}{111}}, \bibinfo{pages}{117001}
  (\bibinfo{year}{2013}).

\bibitem{quijada:1999}
\bibinfo{author}{Quijada, M.~A.} \emph{et~al.}
\newblock \bibinfo{title}{Anisotropy in the \mbox{$ab$}-plane optical
  properties of \mbox{Bi$_2$Sr$_2$CaCu$_2$O$_8$} single-domain crystals}.
\newblock \emph{\bibinfo{journal}{Phys. Rev. B}} \textbf{\bibinfo{volume}{60}},
  \bibinfo{pages}{14917} (\bibinfo{year}{1999}).

\bibitem{puchkov:1996}
\bibinfo{author}{Puchkov, A.~V.}, \bibinfo{author}{Basov, D.~N.} \&
  \bibinfo{author}{Timusk, T.}
\newblock \bibinfo{title}{The pseudogap state in high-\mbox{$T_c$}
  superconductors: an infrared study}.
\newblock \emph{\bibinfo{journal}{J. Phys.: Cond. Matter}}
  \textbf{\bibinfo{volume}{8}}, \bibinfo{pages}{10049} (\bibinfo{year}{1996}).

\bibitem{hwang:2004}
\bibinfo{author}{Hwang, J.}, \bibinfo{author}{Timusk, T.} \&
  \bibinfo{author}{Gu, G.~D.}
\newblock \bibinfo{title}{High-transition-temperature superconductivity in the
  absence of the magnetic-resonance mode}.
\newblock \emph{\bibinfo{journal}{Nature (London)}}
  \textbf{\bibinfo{volume}{427}}, \bibinfo{pages}{714} (\bibinfo{year}{2004}).

\bibitem{yang:2009a}
\bibinfo{author}{Yang, J.} \emph{et~al.}
\newblock \bibinfo{title}{Optical spectroscopy of superconducting
  \mbox{Ba$_{0.55}$K$_{0.45}$Fe$_2$A$_2$}: Evidence for strong coupling to
  low-energy bosons}.
\newblock \emph{\bibinfo{journal}{Phys. Rev. Lett.}}
  \textbf{\bibinfo{volume}{102}}, \bibinfo{pages}{187003}
  (\bibinfo{year}{2009}).

\bibitem{wu:2010b}
\bibinfo{author}{Wu, D.} \emph{et~al.}
\newblock \bibinfo{title}{Eliashberg analysis of optical spectra reveals a
  strong coupling of charge carriers to spin fluctuations in doped
  iron-pnictide \mbox{BaFe$_2$As$_2$} superconductors}.
\newblock \emph{\bibinfo{journal}{Phys. Rev. B}} \textbf{\bibinfo{volume}{82}},
  \bibinfo{pages}{144519} (\bibinfo{year}{2010}).

\bibitem{hwang:2015}
\bibinfo{author}{Hwang, J.}, \bibinfo{author}{Carbotte, J.~P.},
  \bibinfo{author}{Min, B.~H.}, \bibinfo{author}{Kwon, Y.~S.} \&
  \bibinfo{author}{Timusk, T.}
\newblock \bibinfo{title}{Electron-boson spectral density of lifeas obtained
  from optical data}.
\newblock \emph{\bibinfo{journal}{J. Phys.: Condens. Matter}}
  \textbf{\bibinfo{volume}{27}}, \bibinfo{pages}{055701}
  (\bibinfo{year}{2015}).

\bibitem{benfatto:2011}
\bibinfo{author}{Benfatto, L.}, \bibinfo{author}{Cappelluti, E.},
  \bibinfo{author}{Ortenzi, L.} \& \bibinfo{author}{Boeri, L.}
\newblock \bibinfo{title}{Extended drude model and role of interband
  transitions in the midinfrared spectra of pnictides}.
\newblock \emph{\bibinfo{journal}{Phys. Rev. B}} \textbf{\bibinfo{volume}{83}},
  \bibinfo{pages}{224514} (\bibinfo{year}{2011}).

\bibitem{fang:2009}
\bibinfo{author}{Fang, L.} \emph{et~al.}
\newblock \bibinfo{title}{Roles of multiband effects and electron-hole
  asymmetry in the superconductivity and normal-state properties of
  \mbox{Ba(Fe$_{1-x}$Co$_x$)$_2$As$_2$}}.
\newblock \emph{\bibinfo{journal}{Phys. Rev. B}} \textbf{\bibinfo{volume}{80}},
  \bibinfo{pages}{140508} (\bibinfo{year}{2009}).

\bibitem{shen:2011}
\bibinfo{author}{Shen, B.} \emph{et~al.}
\newblock \bibinfo{title}{Transport properties and asymmetric scattering in
  \mbox{Ba$_{1-x}$K$_x$Fe$_2$As$_2$} single crystals}.
\newblock \emph{\bibinfo{journal}{Phys. Rev. B}} \textbf{\bibinfo{volume}{84}},
  \bibinfo{pages}{184512} (\bibinfo{year}{2011}).

\bibitem{marsik:2013}
\bibinfo{author}{Marsik, P.} \emph{et~al.}
\newblock \bibinfo{title}{Low-energy interband transitions in the infrared
  response of \mbox{Ba(Fe$_{1-x}$Co$_x$)$_2$As$_2$}}.
\newblock \emph{\bibinfo{journal}{Phys. Rev. B}} \textbf{\bibinfo{volume}{88}},
  \bibinfo{pages}{180508} (\bibinfo{year}{2013}).

\bibitem{ito:1993}
\bibinfo{author}{Ito, T.}, \bibinfo{author}{Takenaka, K.} \&
  \bibinfo{author}{Uchida, S.}
\newblock \bibinfo{title}{Systematic deviation from \mbox{$T$}-linear behavior
  in the in-plane resistivity of \mbox{YBa$_2$Cu$_3$O$_{7-y}$}: Evidence for
  dominant spin scattering}.
\newblock \emph{\bibinfo{journal}{Phys. Rev. Lett.}}
  \textbf{\bibinfo{volume}{70}}, \bibinfo{pages}{3995} (\bibinfo{year}{1993}).

\bibitem{yi:2011}
\bibinfo{author}{Yi, M.} \emph{et~al.}
\newblock \bibinfo{title}{Symmetry-breaking orbital anisotropy observed for
  detwinned \mbox{Ba(Fe$_{1-x}$Co$_x$)$_2$As$_2$} above the spin density wave
  transition}.
\newblock \emph{\bibinfo{journal}{PNAS}} \textbf{\bibinfo{volume}{108}},
  \bibinfo{pages}{6878} (\bibinfo{year}{2011}).

\bibitem{barisic:2010}
\bibinfo{author}{Barisic, N.} \emph{et~al.}
\newblock \bibinfo{title}{Electrodynamics of electron-doped iron pnictide
  superconductors: Normal-state properties}.
\newblock \emph{\bibinfo{journal}{Phys. Rev. B}} \textbf{\bibinfo{volume}{82}},
  \bibinfo{pages}{054518} (\bibinfo{year}{2010}).

\bibitem{canfield:1992}
\bibinfo{author}{Canfield, P.~C.} \& \bibinfo{author}{Fisk, Z.}
\newblock \bibinfo{title}{Growth of single crystals from metallic fluxes}.
\newblock \emph{\bibinfo{journal}{Phil. Mag.}} \textbf{\bibinfo{volume}{65}},
  \bibinfo{pages}{1117} (\bibinfo{year}{1992}).

\bibitem{wang:2009}
\bibinfo{author}{Wang, X.~F.} \emph{et~al.}
\newblock \bibinfo{title}{Anisotropy in the electrical resistivity and
  susceptibility of superconducting \mbox{BaFe$_2$As$_2$} single crystals}.
\newblock \emph{\bibinfo{journal}{Phys. Rev. Lett.}}
  \textbf{\bibinfo{volume}{102}}, \bibinfo{pages}{117005}
  (\bibinfo{year}{2009}).

\bibitem{ni:2009}
\bibinfo{author}{Ni, N.}
\newblock \bibinfo{title}{Structural/magnetic phase transitions and
  superconductivity in \mbox{Ba(Fe$_{1-x}$TM$_x$)$_2$As$_2$ (TM = Co, Ni, Cu,
  Co/Cu, Rh and Pd)} single crystals}.
\newblock \emph{\bibinfo{journal}{Ph.D. thesis, Iowa State University}}
  (\bibinfo{year}{2009}).

\bibitem{homes:1993}
\bibinfo{author}{Homes, C.~C.}, \bibinfo{author}{Reedyk, M.~A.},
  \bibinfo{author}{Crandles, D.~A.} \& \bibinfo{author}{Timusk, T.}
\newblock \bibinfo{title}{Technique for measuring the reflectance of irregular,
  submillimeter-sized samples}.
\newblock \emph{\bibinfo{journal}{Appl. Opt.}} \textbf{\bibinfo{volume}{32}},
  \bibinfo{pages}{2976} (\bibinfo{year}{1993}).

\bibitem{wooten}
\bibinfo{author}{Wooten, F.}
\newblock \emph{\bibinfo{title}{Optical Properties of Solids}}
  (\bibinfo{publisher}{Academic, New York}, \bibinfo{year}{1972}).
\newblock \bibinfo{note}{(Note: Key material on page 176)}.

\bibitem{tu:2010}
\bibinfo{author}{Tu, J.~J.} \emph{et~al.}
\newblock \bibinfo{title}{Optical properties of the iron arsenic superconductor
  \mbox{BaFe$_{1.85}$Co$_{0.15}$As$_2$}}.
\newblock \emph{\bibinfo{journal}{Phys. Rev. B}} \textbf{\bibinfo{volume}{82}},
  \bibinfo{pages}{174509} (\bibinfo{year}{2010}).

\bibitem{tanner1}
\bibinfo{author}{Tanner, D.~B.} \& \bibinfo{author}{Poter, C.~D.}
\newblock \emph{\bibinfo{journal}{http://www.phys.ufl.edu/~tanner/datan.html}}
  .

\end{thebibliography}


\newpage

\noindent {\bf Author Contribution} J.H. wrote the main manuscript, S.L. took the optical data and analyzed them. K.C. grew the single crystals, E.J. and S.R. contributed to the taking and analyzing of the optical data, S.S. and T.P. took the transport data. All authors reviewed the manuscript.
\\ \\
{\bf Acknowledgements} financial support from the National Research Foundation of Korea (NRFK grant No. 20100008552 and NRFK Grant No. 2013R1A2A2A01067629) and the Basic Science Research Program (2012-008233) funded by the Korean Federation of Science and Technology Societies.
\\ \\
{\bf Competing Interests} The authors declare that they have no competing financial interests.
\\ \\
{\bf Correspondence} Correspondence and requests for materials should be addressed to Jungseek Hwang~(email: jungseek@skku.edu).

\begin{figure}[!htbp]
  \vspace*{-0.3 cm}%
  \centerline{\includegraphics[width=6.0 in]{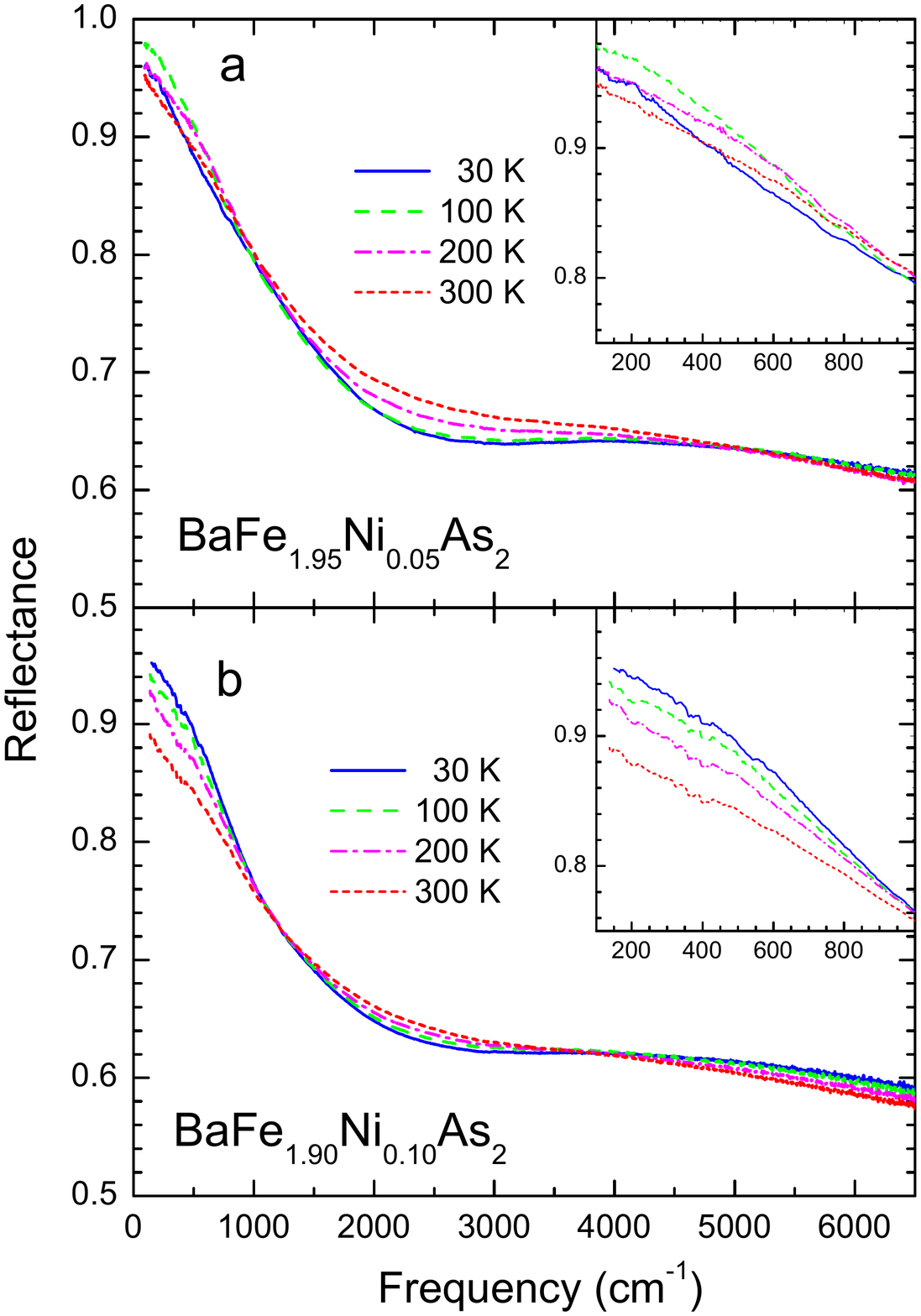}}%
  \vspace*{-0.7 cm}%
\label{fig1}
\end{figure}

\begin{figure}[!htbp]
  \vspace*{-0.3 cm}%
  \centerline{\includegraphics[width=6.0 in]{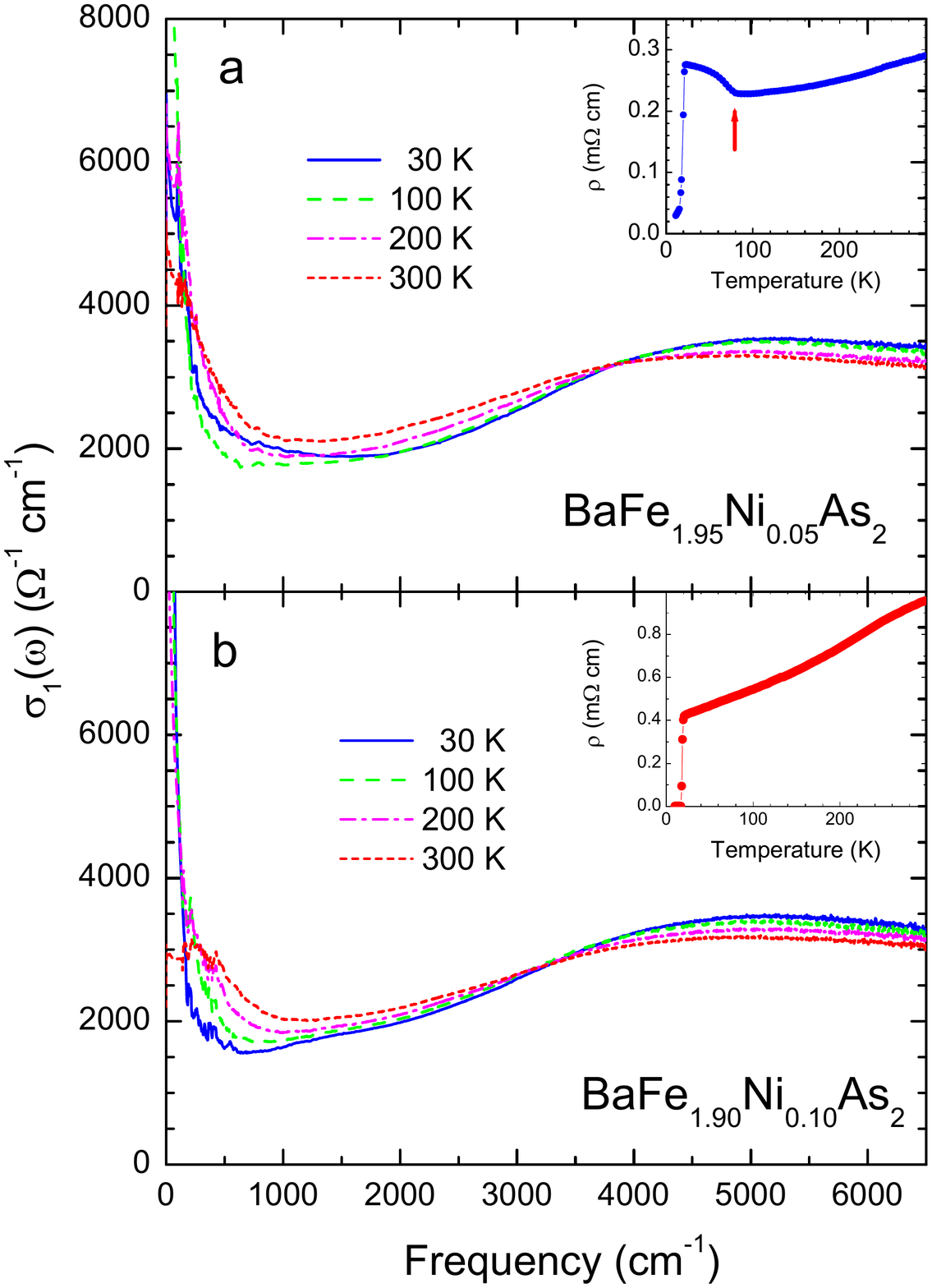}}%
  \vspace*{-0.7 cm}%
 \label{fig2}
\end{figure}

\begin{figure}[!htbp]
  \vspace*{-0.3 cm}%
  \centerline{\includegraphics[width=6.0 in]{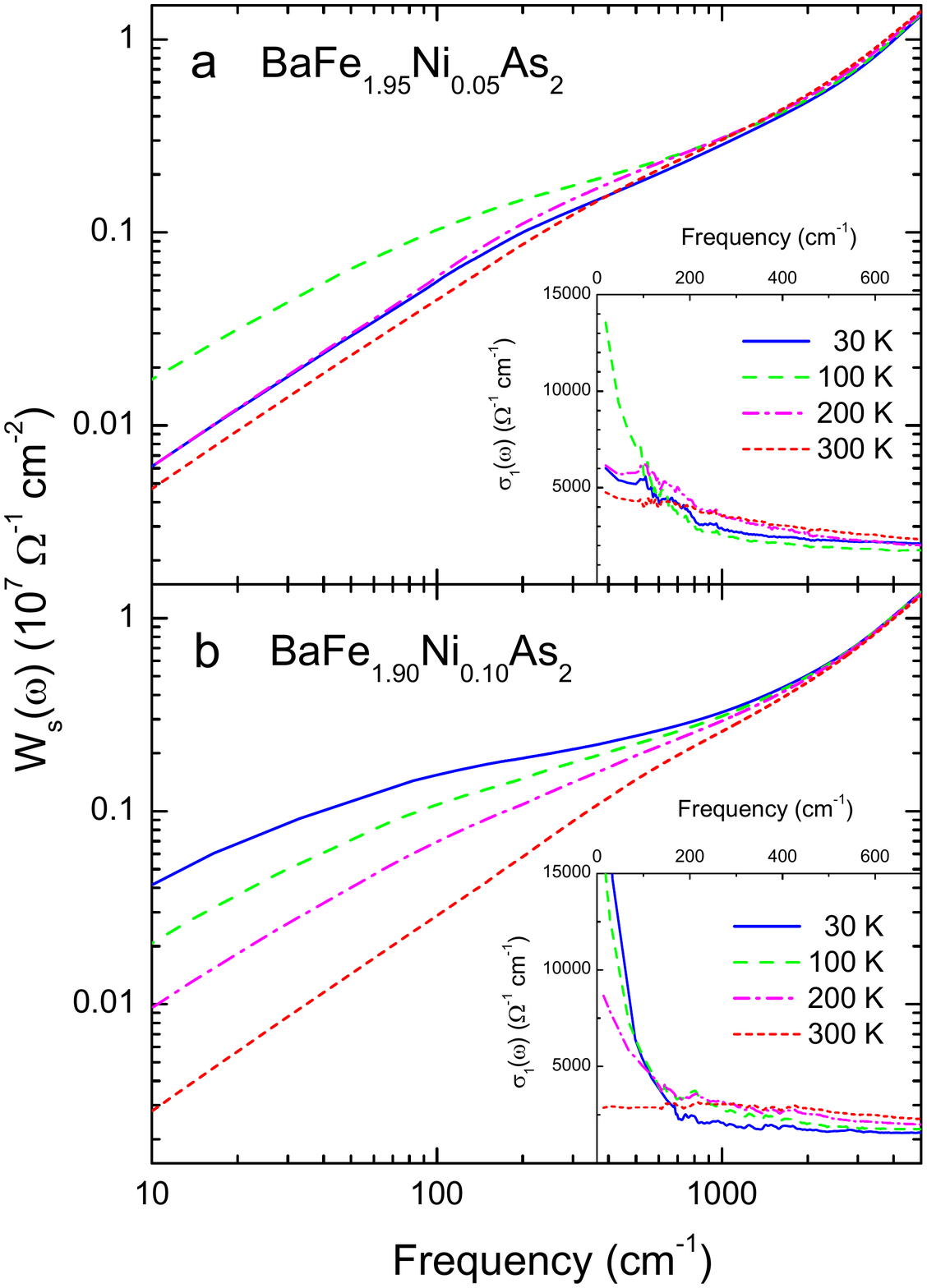}}%
  \vspace*{-0.7 cm}%
 \label{fig3}
\end{figure}

\begin{figure}[!htbp]
  \vspace*{-0.3 cm}%
  \centerline{\includegraphics[width=6.5 in]{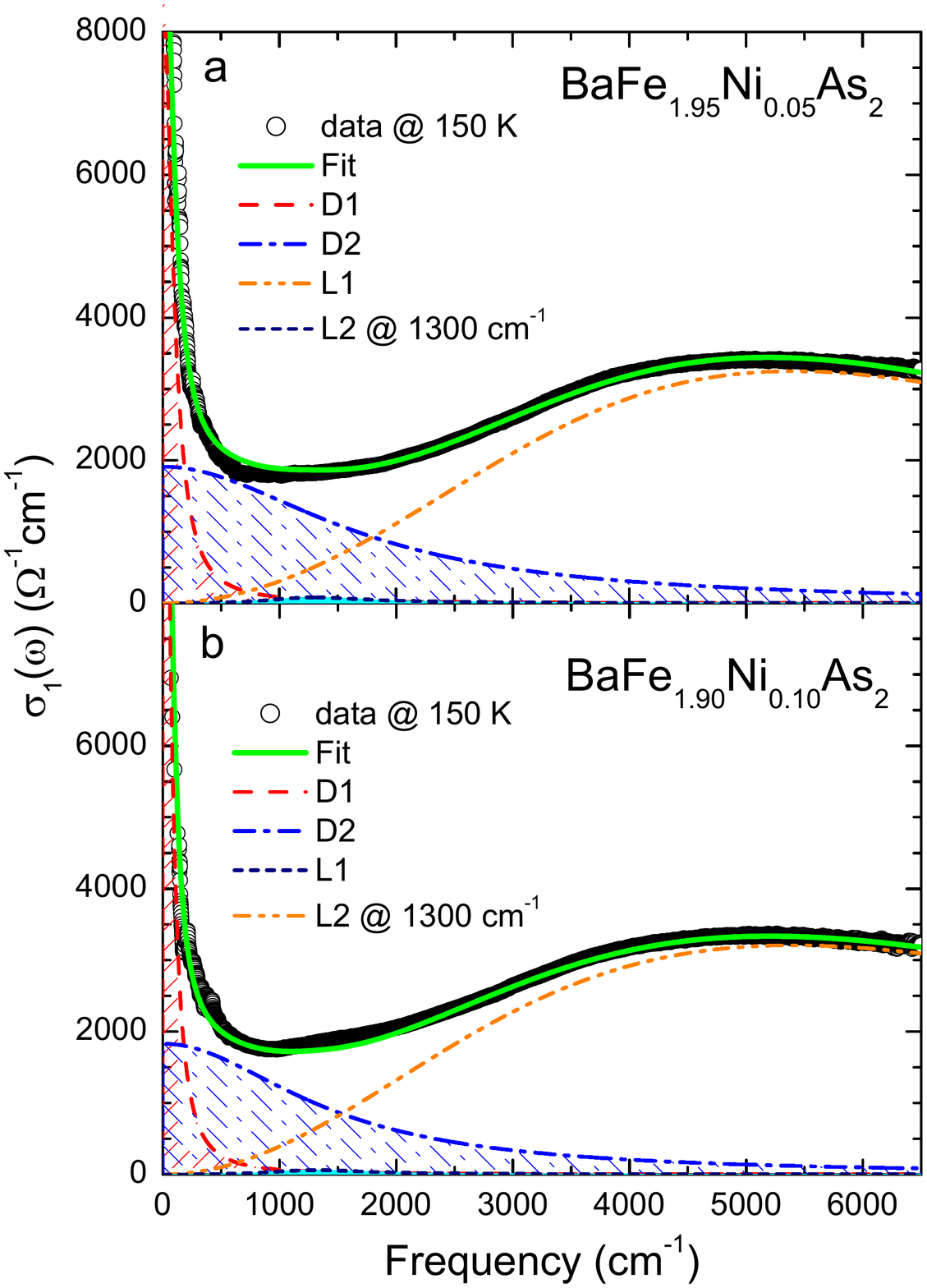}}%
  \vspace*{-0.7 cm}%
 \label{fig4}
\end{figure}

\begin{figure}[!htbp]
  \vspace*{-0.3 cm}%
  \centerline{\includegraphics[width=7.0 in]{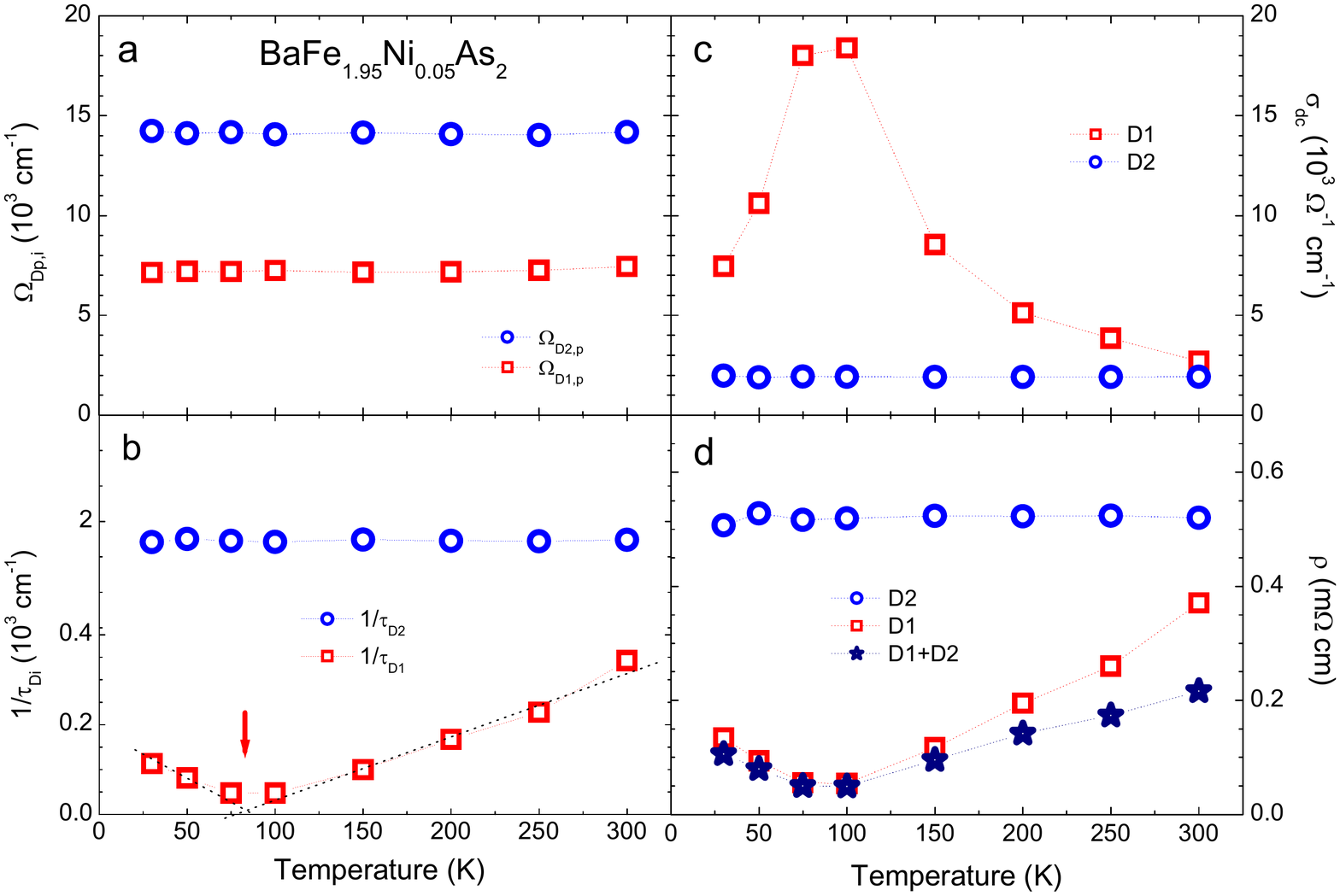}}%
  \vspace*{-0.2 cm}%
 \label{fig5}
\end{figure}

\begin{figure}[!htbp]
  \vspace*{-0.3 cm}%
  \centerline{\includegraphics[width=7.0 in]{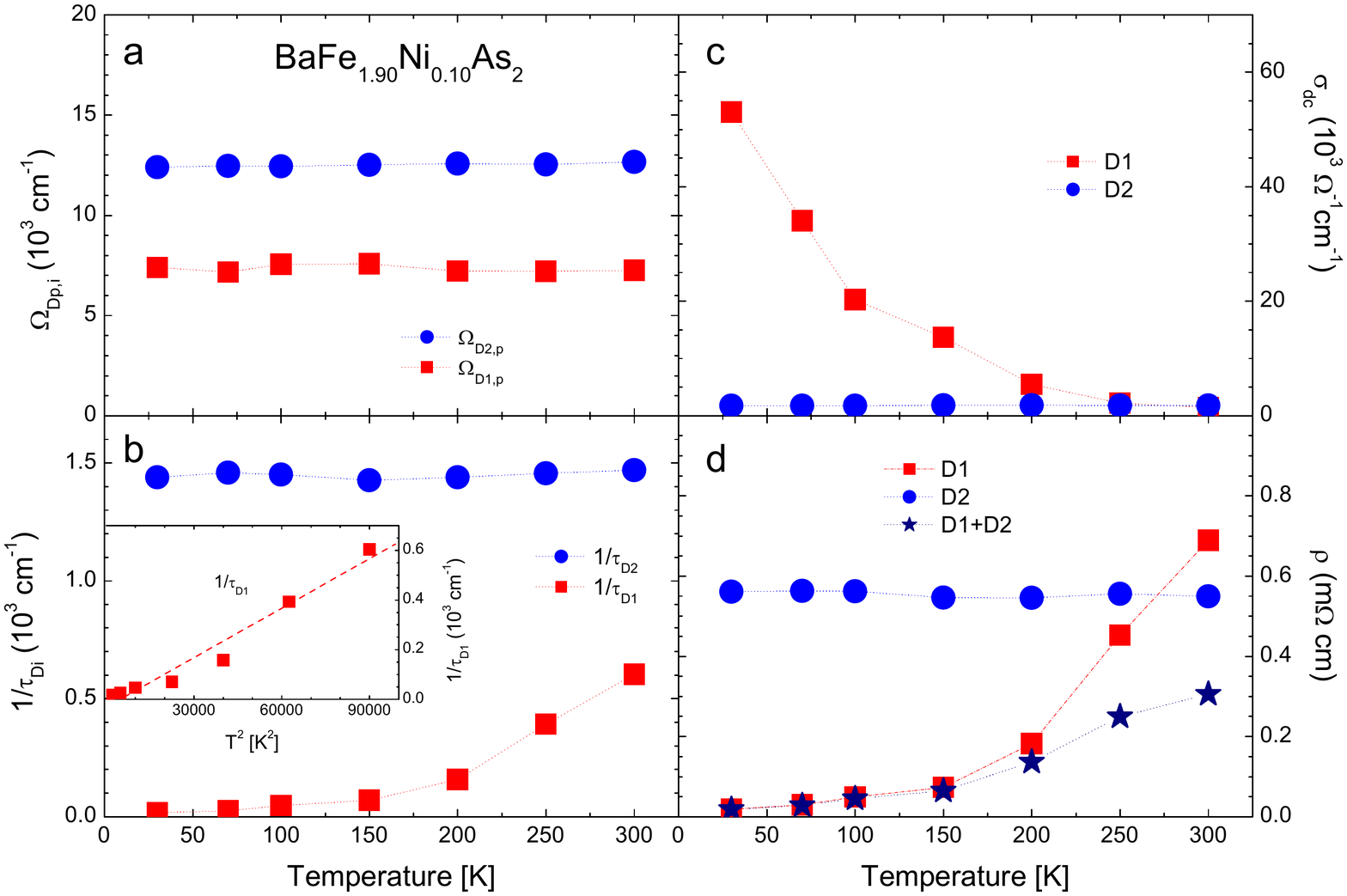}}%
  \vspace*{-0.2 cm}%
 \label{fig6}
\end{figure}


\end{document}